\documentstyle[11pt]{article}
\begin{document}
\twocolumn

\section*{Gas in Galaxies}
\author{Joss Bland-Hawthorn, Ron Reynolds}

The interstellar medium (ISM) can be thought of as the galactic atmosphere
which fills the space between stars.  When clouds within the ISM collapse, 
stars are born. When the stars die, they return their matter to the 
surrounding gas. Therefore the ISM plays a vital role in galactic evolution.

The medium includes starlight, gas, dust, planets, comets, asteroids, fast 
moving charged particles (cosmic rays) and magnetic fields.   The gas
can be further divided into hot, warm and cold components, each of which 
appear to exist over a range of densities, and therefore pressures.
Remarkably, the diverse gas components, cosmic rays, magnetic fields and 
starlight all have very roughly the same energy density of about 1 eV 
cm$^{-3}$.  All the major constituents (or phases) of the interstellar 
medium appear to be identified now, although complete multi-phase studies 
are extremely difficult beyond a few thousand parsecs from the Sun.
The interstellar medium is a highly complex environment which does not 
lend itself to simple analysis. However, this has not stopped astrophysicists
from producing basic models of the ISM in order to make sense of the great 
wealth of data coming in from ground-based telescopes and satellites.

The study of the interstellar medium began around 1927 with the publication of 
Edward Emerson
Barnard's photographic atlas of the Milky Way. The atlas shows dark clouds 
silhouetted against the background star light. 
At about the same time, spectra by John Plaskett and Otto Struve
established the existence of interstellar clouds containing ionized calcium.
By number of nuclei, about 90\% of interstellar matter is hydrogen, 10\% is
helium. All of the elements heavier than helium constitute about 0.1\% of
the interstellar nuclei, or about 2\% by mass. Although roughly half of the 
heavier elements are in the gas phase. Most of
the refractory elements (Si, Ca, Fe) are depleted from the gas phase, and
are locked up in small dust grains mixed in with the gas.  Clouds only
account for about half the mass and 2\% of the interstellar volume. 
A far more pervasive `intercloud' component was not identified until the 
discovery of pulsars and the invention of ultraviolet/x-ray astronomy 
in the mid to late 1960s.

The interstellar medium properties generally depend on the type of 
galaxy, and its distribution shows clear radial trends for a given
galaxy. In disk galaxies, the gas piles up into spiral arms (Fig. 1); this is 
where most of the young stars and supernovae are to be found.

The interstellar medium in galaxies is constantly evolving. Stellar winds
and supernova explosions enrich the gas with heavy elements over the course of
billions of years. In the context of the widely accepted cosmological
model of hierarchical galaxy formation, this may be compensated by the 
accretion of primordial gas in the outer parts of galaxies. Stars are the 
principal source of energy for the ISM. Starlight photons produce 
photoelectric emission from dust grains; these photoelectrons help to heat the 
neutral gas. Ultraviolet photons from the youngest stars ionize atoms and 
dissociate molecules.  The main source of kinetic energy are the supernovae: 
these drive shock waves into the surrounding ISM and are largely responsible 
for its complexity.

\section*{Atomic and Molecular Clouds}

Half of the neutral atomic hydrogen and all of the molecular hydrogen in
the ISM is concentrated into relatively high density and low temperature
regions called `clouds'. The properties of the atomic hydrogen (HI) clouds
have been determined primarily from radio observations of the hyperfine
ground state transitions of hydrogen at 1420 MHz (21 cm); interstellar 
absorption lines of trace elements such as Ca$^+$ also continue to play a
key role in the study of these clouds. For the most common clouds, where 
the 21-cm radiation escapes freely, the brightness of the emission provides
a direct measurement of the HI column density $N_H = \int n_H\ ds$ where
$n_H$ is the atomic hydrogen density. When an HI cloud lies in front 
of a bright source of radio continuum emission,
the decrease in the brightness of the background source at 21 cm is 
proportional to $N_H = \int n_H/T_H\ ds$, where $T_H$ is the temperature
of the HI cloud. Thus, observations of HI clouds at 21 cm in emission and
absorption provide direct information about cloud temperatures and 
column densities.  Table 1 summarises basic properties of the cold and
warm neutral media.  Maps of the sky show that HI clouds have complex shapes
resembling thin extended sheets or filaments with embedded small clumps.

Molecular hydrogen is confined to the interiors of the densest and most
massive clouds, the dark clouds, where starlight capable of dissociating
molecules cannot penetrate. These clouds constitute the active star-forming
component of the interstellar medium. Because H$_2$ has no electric
dipole moment, radiative transitions of H$_2$ are greatly suppressed.
Therefore, most of the structural information about molecular clouds in
the ISM is obtained through observations of the rotational transitions
of the trace molecule CO at 115 GHz (2.6 mm). In addition, a wide variety
of other molecules, including complex hydrocarbon chains, have been
detected within the H$_2$ clouds. Molecular clouds are small (40 pc), dense
(200 cm$^{-3}$), with structure on scales of less than 0.1 pc (see Table 1). 
Some of the small condensations can have densities as high as 10$^5$ cm$^{-3}$.
In disk galaxies like our own, the cold neutral and molecular gas are confined 
to a disk which is much thinner than the stellar disk.

\section*{The Intercloud Medium}

Astrophysicists have long wondered as to what confines gas clouds within the
Galaxy. In 1956, Spitzer
speculated that a rarefied ($\sim 10^{-3}$ cm$^{-3}$), hot gas ($\sim 10^6$ K),
extending a kiloparsec or more above the galactic plane, would confine the 
diffuse gas clouds observed far above the plane and would prevent their 
expansion and dissipation.  Confirmation of the predicted corona came 17 years 
later with the advent of space astronomy (Table 1). Present day satellites 
allow for the direct detection of diffuse x-ray emission and ultraviolet/x-ray 
spectral lines from the highly ionized trace elements within the hot gas
(see the review by Spitzer 1990). The hot gas is thought to arise from the
action of supernovae as we discuss below (see Fig. 4).

The existence of widespread hot gas in the disk of the Galaxy comes from
observations of O$^{+5}$ ions at ultraviolet wavelengths, and the direct
detection of soft x-ray emission. A clear demonstration of coronal halo gas 
far from the disk has been harder to come by. A wide range of ions is
observed (Si$^{+3}$, C$^{+3}$, N$^{+4}$) towards the halo. Current models
suggest that ultraviolet light from disk stars can only account for some 
of the ionization. The N~V absorption lines appear to require collisional 
heating from a pervasive hot corona (Sembach \& Savage 1992).

From a theoretical standpoint, the expectation is that hot young stars and
supernovae punch holes or blow bubbles (Fig. 2) in the
surrounding gas, and the diffuse hot component escapes into the halo 
through buoyancy. In fact, there are spectacular examples of bubbles 
seen at 21 cm in the Galactic interstellar medium (see Fig. 3).
But there are many outstanding problems with these models, not least of 
which are complications imposed by the magnetic field. Since the gas is 
thermally unstable, it is equally probable that the gas undergoes a 
`cooling flow' or `fountain flow' back towards the disk. 

Filling the space between clouds are two additional components of the 
intercloud medium. By mass, most of the intercloud medium is in the
form a `warm neutral' or a `warm ionized' medium. These phases extend
far beyond the thin disk of cold gas (Table 1). The existence of
the warm ionized medium, was firmly 
established by 1973 from three independent observations: (i) low
frequency radio observations by Hoyle \& Ellis, (ii) time delays in radio
pulses from pulsars (see below), and (iii) through direct observation
of H$^+$ recombination emission by wide-field Fabry-Perot interferometers.
This gas has a density of roughly 0.1 cm$^{-3}$ and a temperature near 10$^4$K.
The dominant source of ionization appears to be dilute ionizing flux from
young stars in the disk, although some models suggest that the cooling
radiation in old supernova remnants can be important.
The deepest optical spectra to date show that the ionized
gas extends to at least 5 kpc into the halo in some cases, and
extends even further in radius than the HI disk.
 
Roughly half of the interstellar HI appears to be located in the `warm
neutral' component of the intercloud medium. This intercloud HI was first
identified in 1965  as the source of the ubiquitous, relatively broad
(velocity dispersion $\sim 9$ km s$^{-1}$) 21-cm emission features that
had no corresponding absorption when viewed against bright background
radio sources. The large velocity dispersions and the absence of absorption
imply temperatures of 5000 to 10,000 K. Observations of the Ly$\alpha$ 
absorption line of HI toward bright stars show that this gas has a mean
extent from the midplace of 500 pc, i.e., much thicker than the cold
neutral disk. If the warm neutral medium is in pressure equilibrium with
the cold component, then it would be clumped into regions occupying 35\%
of the intercloud volume with a density of 0.3 cm$^{-3}$ at the midplane
(Table 1), although these numbers are highly uncertain.

\section*{The Solar Neighborhood}

We now consider the interplanetary medium (heliosphere) within the Solar 
System as distinct from the local interstellar medium. The Sun moves
with a velocity of about 20 km s$^{-1}$ relative to the local interstellar
medium. The solar wind produces a bow shock ahead of the Sun. This 
discontinuity (heliopause) defines the extent of the heliosphere in
the upstream direction. It is anticipated that the Pioneer 11 or Voyager 
deep space probes will eventually confirm the existence of this boundary.
The neutral interstellar gas is largely unaffected by the heliopause,
so that the local neutral medium presumably streams relatively
freely through the Solar System. In contrast, ions and charged dust grains
are probably deflected by the advancing heliosphere. Cosmic rays 
are relativistic particles and therefore penetrate the heliosphere. 

The most local gas we can associate with the local interstellar medium 
has the properties of the warm neutral medium. The solar radiation 
is backscattered by the local medium. Studies which exploit the resonance
lines of hydrogen and helium show that the gas has a temperature of 
8000 K, a hydrogen density of 0.25 cm$^{-3}$ and a helium density of
0.02 cm$^{-3}$. Absorption line measurements toward nearby stars indicate
that this warm gas is only a few parsecs in extent with hotter 10$^6$K
gas occupying most of the volume within 100~pc of the Sun.

\section*{Theories of the ISM}

\subsection*{The two-phase model}

Field, Goldsmith \& Habing (1969) developed the first quasi-static theory
of a multi-phase ISM. The FGH model contained two stable phases, the `cold'
neutral phase at a temperature of $\sim$100K and a `warm' phase at 
$\sim 10^4$K where about 10\% of the gas is ionized. The model uses cosmic 
ray heating to balance the cooling of these two phases. The cooling is
dominated by fine-structure excitations from C$^+$ in the cool gas, 
and collisional excitations (Ly$\alpha$) in the warm gas. The model predicts 
an ISM where the majority of the volume is occupied by the warm ($\sim 10^4$K) 
intercloud medium while the majority of the mass is contained within the dense 
cold clouds.

However, the FGH model indicated that the ISM should be stratified 
perpendicular to the Galactic disk, which was known to violate the 
observed properties of the ISM. They therefore appealed to the collective
explosive effects of HII regions and supernovae, which impart sufficient
turbulent motions to destroy any gravitationally induced stratification.
The FGH model allows the cold clouds to have a large range of sizes. The
smallest clouds are destroyed by thermal conduction within the warm layer,
while the largest are gravitationally unstable and prone to collapse.
Otherwise, there is no limit on the size or shape of the expected neutral
structures. 

FGH also raised the possible existence of another stable phase at
a temperature of $\sim 10^6$K, dominated by bremsstrahlung cooling. 
However, prior to observations of the soft x-ray background or high-ionization
UV spectra, there was no need to invoke the actual existence of this medium.

In the FGH model, supernovae (and HII regions) were required to destroy the
gravitationally-induced vertical stratification of the ISM. This inspired
theoretical models of the expansion and thermal evolution of supernova
remnants, and it was then that the full importance of supernovae came to
be realized. Supernovae generate blast waves with speeds of 10,000 km s$^{-1}$
which shock the ambient
material, producing gas at $\sim 10^6K$ which then slowly cools through
bremsstrahlung emission and line emission from highly ionized ions. 
The remnant structure is a hot bubble surrounded by a thin, relatively dense 
shell of cool $\sim 10^3$K gas. This situation proves to
be quite stable; the hot bubble cannot expand through the shell expanding
ahead of it. After a million years, the radius of the hot bubble is roughly
80 pc. At this point, the interior pressure of the bubble reaches the
ambient pressure of the ISM. This static situation remains until the 
hot bubble cools and contracts after four million years or so.

Cox \& Smith (1974) were the first to recognize the importance of this
evolutionary sequence for the dynamical and thermal balance of the ISM.
If an expanding supernova remnant happens to intersect the static hot
cavity of another remnant, the expanding remnant `breaks out' through
the old, hot remnant. Due to the low density within the static bubble,
the expanding shock wave propagates preferentially into it, reheating
the matter inside. Depending upon the rate of Galactic SNR, the ISM
could therefore evolve towards an interconnected tunnel network filled
predominantly by hot gas.

To estimate the importance of such a tunnel network, a porosity parameter
was introduced such that
\begin{equation}
q = r \tau V_{sn} 
\end{equation}
where $r$ is the average SN rate per unit volume, $\tau$ is the lifetime
of an isolated bubble, and $V_{sn}$ is the final volume of the average
SN-generated bubble. For $q \ll 1$, $q$ is the proportion of the total
interstellar volume that would be filled with hot bubbles. For a 
SN frequency of 1 per 80 yr, they found $q\sim 0.1$. This implies that 
10\% of interstellar space should be filled by unconnected, hot, SN-generated
bubbles. Therefore, the probability of any new SN occurring within a 
preexisting cavity was 0.1, but the probability that any new SNR would
expand into and reheat another older remnant was 0.55.
The consequences of the overlap on the phase structure of the ISM were
radical.

This chain of reasoning rests heavily upon supernovae occurring randomly
throughout the Galactic volume and the expansion of the hot bubble not
being inhibited by magnetic pressure, for example. The simple conclusion 
prompted many 
to believe in the `porosity imperative', that is the need for a hot
$10^6$K phase to occupy the majority of the interstellar volume. The cold
and warm ($\sim 10^4$K) phases were relegated to the walls between the
pervasive bubbles. Large filamentary structures in the ISM seem to support
this, except that the overall pervasiveness and smoothness of 
the ISM is in conflict with this picture.

\subsection*{The three-phase model}

This paved the way for a true theoretical tour de force, the McKee-Ostriker
(MO) model of the supernova-dominated, three-phase ISM. In their model, 
supernovae produce 
the `hot ionized medium' (HIM), the $\sim 10^6$K component of the ISM in 
their bubble interiors, as well as enhancing the formation of the 
`cold neutral' (CNM), `warm neutral' (WNM) and `warm ionized' (WIM) media 
along the compressed edges of remnants. There are still only two fundamental
phases (CNM, HIM) in the MO theory. The WNM and WIM are restricted to the 
interface regions of the neutral clouds, and the WIM in direct contact with 
the HIM and photoionized by thermal emission from it. 

The model attempts to balance the thermal and mass exchange between the 
different phases. The energy input from supernovae is offset by radiation 
cooling 
from the four media.  The mass lost by cloud evaporation into the HIM is 
balanced by dense shell formation of swept up interstellar material. The 
model is known to be incorrect in many details.  There are numerous 
observations of local, dense clouds that are highly overpressured with 
respect to their environment. The model does not treat the influence of 
magnetic fields which are known to thread throughout the ISM and are likely 
to suppress thermal evaporation.  The magnetic fields are of the right
strength to play an important role in equilibrating the pressure balance.
But in the absence of a better working model, the MO theory remains the 
dominant conceptual framework. 

It remains unclear whether thermal pressure balance between the phases is an 
essential feature of the MO picture.
Pressure equilibrium should exist between the cosmic rays,
the magnetic field and the kinetic interstellar component. But the latter need
not be the thermal pressure of the gas. Random, turbulent `bulk' motions of the
gas have a high enough energy density to provide the required effective 
pressure.  The minimum interstellar pressure, $P = n T$, is probably about 
$\sim$25,000 cm$^{-3}$ K and may even be higher.

\section*{The Influence of Supernovae}

After a burst of star formation has taken place, the first supernovae appear 
after about 10 million years. These Type II supernovae arise from collapsing 
massive stars which have exhausted most of their nuclear fuel. Type II 
supernovae are distributed like the young stars in a disk galaxy, i.e. 
concentrated along spiral arms.
A billion years after the initial starburst, the Type I supernovae appear.
Type Ia supernovae are thought to be due to accretion of matter by a white
dwarf in a binary pair. These are to be found among the older stellar 
population and therefore have an exponential distribution with radius like the 
underlying disk.

The increase in Type Ia supernovae towards the Galactic centre implies that 
here the ISM likely
becomes supernova-dominated, resulting in a two phase ISM with only an
HIM and a CNM. The warm phases are disrupted as diffuse clouds are shocked and
heated to high temperatures. The molecular clouds survive the assault of 
supernovae and remain essntially intact, having only their outer warm layers
stripped away. The scale height of the oldest stellar populations is higher
($\sim 300$ pc) than for the atomic hydrogen ($\sim 100$ pc), such that
half of all Type Ia supernovae explode above the gas and therefore deposit much
of their energy directly to the halo. This picture produces a volume filling 
fraction of about 25\% for the HIM, and presumably explains the HI holes seen 
in external galaxies. 

The correlated distribution of Type II supernovae has an important consequence.
The time interval between successive supernovae is less than the bubble lifetime.
This can result in a large-scale wind of energy into the Galactic halo. Heiles
(1987) derives a two-dimensional porosity parameter
\begin{equation}
q_{2D} = 8.3 \sigma s N^{-1}
\end{equation}
where $\sigma$ is the supernova rate in kpc$^{-2}$ Myr$^{-1}$, $s$ ($\sim 2$) 
is a correlation factor, and $N$ ($\sim 40$) is the number of Type II 
supernovae that 
occur in a single association. But this leads to a volume filling factor of more
than 95\% along the spiral arms, and about 80\% outside of the arms, contrary
to observation. Furthermore, the Type II energy flow is expected to break out
and produce a mass flow rate of $\sim$ 20 M$_\odot$ per year, which would 
deplete the total gas content of the disk in 10$^9$ yr. How are we to 
reconcile this?

\section*{The Role of Pulsars}

Discovered in 1967, pulsars are the rapidly rotating core left behind by
exploding Type II supernovae.  More than a thousand have now been observed in 
our Galaxy, in halo globular clusters, and in the LMC and SMC.
Radio observations of these rapidly pulsating stars provide information on the
plasma densities and magnetic field strengths of 
the intervening interstellar medium over base lines of tens of kiloparsecs.

The electrodynamic properties of a plasma are functions of the frequency of
the electromagnetic wave that traverses the plasma. The group velocity of
a wave in a plasma is 
\begin{equation}
v_g = c \left(1-{{\omega^2_p}\over{\omega^2}}\right)^{1/2}
\end{equation}
where $\omega$ is the angular frequency and
$\omega_p$ is the plasma frequency defined by $\omega^2_p= 4\pi n e^2/m$,
for which $n$, $e$ and $m$ are the number density, charge and mass of the 
free electrons.

Because the group velocity of a wave depends on its frequency, the fourier
components of a pulse will traverse a total distance $d$ through a plasma
in a time $t_\omega$ given by
\begin{equation}
t_\omega = \int_0^d {{ds}\over{v_g}}
\end{equation}
where $s$ defines a small increment of distance through the plasma.
Plasma frequencies in interstellar space are typically very low so 
we can expand eqn.~3 and substitute into eqn.~5 such that the time taken for
the fourier component of a pulse to traverse a plasma is
\begin{equation}
t_{\omega} \approx {{d}\over{c}} + (2 c \omega)^{-1} \int^d_0 \omega^2_p\ ds.
\end{equation}
The first term is the time taken to traverse a distance $d$ in vacuo, and
the second term is the plasma correction.

Studies of the arrival times of the various fourier components indicate that
the highest frequencies arrive ahead of the low frequency components. What 
is actually measured is the derivative of eqn.~5,
\begin{equation}
{{dt_\omega}\over{d\omega}} = -{{4 \pi e^2}\over{c m \omega^3}} D_m.
\end{equation}
The dispersion measure, defined as $D_m = \int_0^d n_e\ ds$, 
is a measure of the total column of free electrons along the path to
the pulsar in units of pc cm$^{-3}$. A column of 10$^{20}$ electrons results
in a $D_m$ of 30 pc cm$^{-3}$ and a delay of 12 sec for a signal at 100 MHz
relative to infinite frequency.  Dispersion measures were first obtained by 
the pulsar discovery team. Since then, we have come to learn (Fig.~5)
from pulsars in distant globular clusters and  
the Magellanic Clouds that the warm atmosphere in the Galaxy 
extends to about a kiloparsec above the Galactic plane.

The radio signals reveal other important properties about the diffuse 
gas. In the presence of a magnetic field along the path to the pulsar,
$B_\Vert$, the plane of polarization of the propagating wave will rotate
by an angle $\chi$ equal to the phase delay between the ordinary and
extraordinary components of the electric field. The so-called Faraday 
rotation angle is given by
\begin{equation}
\chi = {{\pi}\over{c \omega^2}} \int_o^d \omega_p^2 \omega_B\ ds
\end{equation}
for which $\omega_B$ is the cyclotron frequency.
We define a quantity called the rotation measure,
\begin{equation}
R_m = 0.81 \int_o^d n_e B_\Vert\ ds
\end{equation}
in the traditional units of rad m$^{-2}$, and
where $B_\Vert$ is in units of microgauss ($\mu$G). The ratio of $R_m$ to
$D_m$ is the average galactic magnetic field strength along the path,
\begin{equation}
\langle B_\Vert \rangle = 1.232\ {R_m \over D_m} .
\end{equation}

\section*{The Influence of Magnetic Fields}

Rotation measures towards 500 pulsars show that there is a random and ordered
magnetic field dispersed throughout the Galaxy. The ordered component has a 
field strength of $1-2$ $\mu$G, while the random component is $5-6\ \mu$G with
a cell size of about 50 pc.

The random field has a dramatic effect on the evolution of supernovae. While it
can be neglected in the early stages of the explosion, as the remnant expands,
the trapped magnetic field in the thin, cool shell resists the expansion, and
results in a much thicker shell with much lower compression. Now, after a 
million 
years, the hot bubble has an inner edge at 60 pc and an outer edge of 90 
pc. Over the next four million years, the combined thermal and magnetic
pressure of the thick shell forces the hot bubble to {\it decrease} in size.
After $\sim 5\times 10^6$ yr, the bubble radius is only 10 pc or so.
Supernova bubbles can still overlap, but generally at a time when they are
significantly weaker disturbances. This suggests that, consistent with the
observations, the HIM produced by Type II supernovae should be much smaller 
than predicted by Heiles.

\section*{Dependence on Galaxy Type}

Surveys of several hundred galaxies show a systematic trend in the
fraction of molecular gas to neutral gas for galaxies along the 
`Hubble sequence' (see Fig.~6). This relationship is not fully 
understood although possible answers include the effect of the
large-scale gravitational field in the formation or disruption
of molecular clouds. For example, the stronger gravitational field 
in the bulge-dominated (early) galaxies may encourage fragmentation
of the gas as a first step to forming dense clouds through collapse.
It appears that there is also an enhanced fraction
of molecules to neutral atoms in merging galaxies and cluster
galaxies. In the latter case (e.g. Virgo), it is
thought that the diffuse hydrogen has been swept away by the
intracluster gas as the galaxy moves through the cluster.

\section*{Unanswered Questions}

On the subject of the interstellar medium, there are vastly more questions than 
answers. There are few topics in astrophysics which are so well served by
the new and planned generation of space-borne and ground-based observatories.
These highly technological and expensive facilities provide the necessary 
impetus to encourage progress from theorists and computational analysts alike.

Although the principal components of the interstellar medium have been
identified and many of their properties measured, there is very little
understanding of how they fit together into a dynamic system. This lack of
progress has begun to stifle progress in other astrophysical fields.
What are the feedback mechanisms that determine the stellar and interstellar
properties of a galaxy? What regulates the star formation rate? Why do
the many components of the ISM 
have very roughly the same energy density?
What is the volume filling fraction and topology of the various gas phases?
What is the large-scale and fine-scale topology of the magnetic field for
each of the gas phases?
How do the properties of the ISM change with cosmic time as more and more
of the gas becomes locked up in stars?

To quote from Spitzer (1990), ``Understanding the processes that occur as
the hot interstellar gas evolves in our Galaxy is an ambitious goal that
we are far from achieving.'' The dynamics of a compressible gas bombarded
by photons and cosmic rays is a highly complex problem. Progress has been
made through idealized models. In these models, one commonly recognizes
three phases. Initially, the supernova ejects a rapidly expanding 
envelope whose interaction heats the ambient gas to high temperatures.
Second, the hot gas expands and compresses or destroys clouds in its wake. 
Finally, the hot gas may escape or fall back down to the galactic plane.
The problem is that, in practice, these three stages overlap so much that
their mutual interactions are crucial. Since we can expect major gains in 
the computational power of supercomputers, we can anticipate the development
of more realistic models in the decades ahead.

\section*{Bibliography}

\noindent
 Cox D P and Smith B W 1974 Large-scale effects of supernova remnants on the 
Galaxy: generation and maintenance of a hot network of tunnels {\it Astrophys.
J.} {\bf 189} L105--L108

\medskip\noindent
 Field G B, Goldsmith D W and Habing H J 1969 Cosmic ray heating of the 
interstellar gas {\it Astrophys. J.} {\bf 155} L149--L154

\medskip\noindent
 Heiles C 1987 Supernovae versus models of the interstellar medium and the 
gaseous halo {\it Astrophys. J.} {\bf 315} 555--566

\medskip\noindent
 Mac~Low, M-M, McCray, R and Norman, M L  1989, {\it Astrophys. J.}

\medskip\noindent
 McKee C F and Ostriker J P 1977 A theory of the interstellar medium - three 
components regulated by supernova explosions in an inhomogeneous substrate
{\it Astrophys. J.} {\bf 218} 148--169

\medskip\noindent
 Sembach K and Savage B 1992 Observations of highly ionized gas in the Galactic 
halo {\it Astrophys. J. Suppl.} {\bf 83} 147--201

\medskip\noindent
 Spitzer L 1990 Theories of the hot interstellar gas {\it Ann. Rev. Astron. 
Astrophys.} {\bf 28} 71--102

\vskip 2cm
\noindent Table 1: The component properties of the interstellar medium

\begin{tabular}{lcccc} \hline
component & temperature & midplane            & filling       & average \\
          &   (K)       & density (cm$^{-3}$) & fraction (\%) & height (pc) \\
\hline
 & & & & \\
{\it Clouds} & & & & \\
H$_2$ & 15 & 200 & 0.1 & 75 \\
HI    & 120 & 25 & 2 & 100 \\
 & & & & \\
{\it Intercloud} & & & & \\
Warm HI & 8000 & 0.3$^*$ & 35$^*$ & 500 \\
Warm HII & 8000 & 0.15 & 20 & 1000 \\
Hot HII & $\sim10^6$ & 0.002 & 43$^*$ & 3000$^*$ \\
 & & & & \\
\hline
\end{tabular}

\noindent
$^*$ Value uncertain by at least a factor of 2.

\vskip 2cm
\section*{Figure captions}

\medskip\noindent {\bf Fig. 1.} 
A colour composite of the northern spiral arm 
in M83. The molecular gas is shown in blue, the 20 cm radio continuum in 
red, and the ionized gas in green (Courtesy of R. Rand, University of 
New Mexico). Note that the cold gas, warm gas and dust pile up into spiral
arms.  Young stars form within the cold gas and then warm up the gas and dust
through photoionization. Eventually, the young stars evolve to become
supernovae which interact violently with the gas and dust (see Fig.~2).
The rotation of the gas and stars is clockwise such that the spiral arms
trail and the stars and gas overtake the spiral arms from the concave side.

\medskip\noindent {\bf Fig. 2.} A simulation of the breakout from a galactic 
disk of a superbubble driven by multiple supernovae.  The density of gas in 
the disk and superbubble are shown in cross-section, with only the upper half 
of the disk shown.  Red is high-density gas, while blue is low-density gas, 
with other colors of the rainbow intermediate.  Each side of the image is 
approximately 800 pc long.  (Courtesy of Mordecai-Mark Mac Low, American 
Museum of Natural History).

\medskip\noindent {\bf Fig. 3.} 
An image of a large Galactic HI supershell (white region at center).  
The empty supershell has a central brightness temperature of about 3~K;
the shell edges have a brightness temperature around 60~K (black).  
The shell also shows narrow channels
which appear to extend to the Galactic halo, forming a "chimney" above and
below the plane.  The shell lies at a distance of about 6.5~kpc, has a
diameter of roughly 600 pc and extends more than 1.1 kpc above the Galactic
plane.  The data were obtained at the Parkes Radio Telescope as part of
the Southern Galactic Plane Survey.
(Courtesy of N.M. McLure-Griffiths \& J.R. Dickey, University of Minnesota).

\medskip\noindent {\bf Fig. 4.} 
These images show ROSAT false-color images of the Corona Australis dark 
molecular cloud. The contours show the 100 micron emission from dust in this 
cloud measured by the IRAS infrared satellite. The self-scaled images are 
for two different energy bands: (a) 100 eV $-$ 300 eV (C band),  and (b) 
500 $-$ 1100 eV (M band). 
These soft (low energy) x-rays are absorbed by interstellar dust and 
gas. Because the M band x-rays are more penetrating than the C band
x-rays, they are absorbed more strongly in the core of the cloud than in the 
periphery, while the C band x-rays from beyond the cloud are completely 
absorbed over the entire cloud. These images demonstrate that much of the
x-ray flux originates from beyond the cloud.
(Courtesy of D.N. Burrows, Penn State University.)

\medskip\noindent {\bf Fig. 5.} 
Dispersion measures from pulsars, $D_m \sin\ b$ where $b$ is galactic
latitude, versus $z-$distance above the Galactic plane. The horizontal
lines show the distance uncertainty for different pulsars. The black
circles and stars refer to pulsars in globular clusters and the Magellanic
clouds respectively. The sloping line corresponds to a model electron
distribution which is uniform in density 0.03 cm$^{-3}$, and the two
dashed lines are for models in which the electron layer has the same
density at $z=0$ but falls off with increasing $z$ as sech$^2(z/h)$
where $h$ is 500~pc and 800~pc.
(Courtesy of R.N. Manchester, Australia Telescope National Facility.)

\medskip\noindent {\bf Fig. 6.}
The ratio of molecular to atomic hydrogen for different `Hubble types' of 
spiral galaxies. The histograms are presented in order of declining 
bulge to disk ratios, with the S0/Sa galaxies having the largest bulges.
The increasing fraction of dense molecular clouds 
towards earlier Hubble types may reflect a higher cloud formation
rate in the presence of a stronger gravitational field.
(Courtesy of J.S. Young, University of Massachussetts.)

\vskip 3cm
\noindent {\sl Joss Bland-Hawthorn \& Ron Reynolds}

\end{document}